\documentclass[journal]{IEEEtran} 
\usepackage{graphicx}
\usepackage{color}
\usepackage{blindtext}
\usepackage{eso-pic}
\usepackage{amsmath}
\usepackage{gensymb}
\usepackage{authblk}
\usepackage{fancyref}
\usepackage{tikz}
\usepackage[justification=centering]{caption}

\newcommand{\sn}[2]{\ensuremath{{#1}\times 10^{#2}}} 

\newcommand\AtPageUpperMyright[1]{\AtPageUpperLeft{%
 \put(0,\LenToUnit{-1cm}){%
     \parbox{0.66\textwidth}{\raggedleft\fontsize{10}{12}\selectfont #1}}%
 }}%
\newcommand{\conf}[1]{%
\AddToShipoutPictureBG*{%
\AtPageUpperMyright{#1}
}}

\newcommand*{\TitleFont}{%
      \usefont{\encodingdefault}{\rmdefault}{b}{n}%
      \fontsize{23}{20}%
      \selectfont}
			
\begin{document}
\title{\TitleFont Safety-Interlock System of the DSSC X-Ray Imager}

\author[1,2]{Sneha~Nidhi~\IEEEmembership{Member,~IEEE,}}
\author[1]{Helmut~Klaer}
\author{Karsten~Hansen~\IEEEmembership{Senior Member,~IEEE,}}
\author[2]{Monica~Turcato}
\author[2]{Markus~Kuster}

\author[2]{Matteo~Porro~\IEEEmembership{Member,~IEEE,}}
\affil[1]{Deutsches Elektronen-Synchrotron DESY, Hamburg, Germany}
\affil[2]{European XFEL, Schenefeld, Germany}

\maketitle
\conf{2016 IEEE Nuclear Science Symposium Conference Record (NSS/MIC)}

\pagestyle{empty}
\thispagestyle{empty}

\begin{abstract}
In this paper, the DSSC safety-interlock system is introduced. It is designed to keep the DSSC mega-pixel camera in a safe-state. The system is composed of four inter-communicating sub-systems referred here as a master SIB (safety-interlock board) and three partner SIBs. Each SIB monitors and processes 75 temperature sensors mainly located in the focal plane and also distributed inside the camera-head electronics. It monitors the signals from two pressure sensors, one humidity sensor as well as the connectivity to sixteen low-voltage and two high-voltage cables. The master SIB supervises the signals to and from external equipments like the cooling system, pressure gauge, power crates and status information from experimental environment. SIB is an 8-layer printed-circuit board (PCB) with a micro-controller sitting on top as its central processing unit. A decision-matrix software reads the sensor status continuously and determines whether the camera is in a safe state or not. All error and warnings flags are sent to the detector control software framework and the end user directly. These flags are also stored every 100 ms together with the image data. The full SIB sensor data is sent to the back end for offline analysis. A shutdown sequence is initiated by SIB in case of a critical failure. The safety-interlock philosophy, hardware design, and software architecture along with test results will be presented. 
            
\end{abstract}


\section{Introduction}

\begin{figure}[htbp]
\centering
\includegraphics[width=.49\textwidth,trim=0 0 0 0,clip]{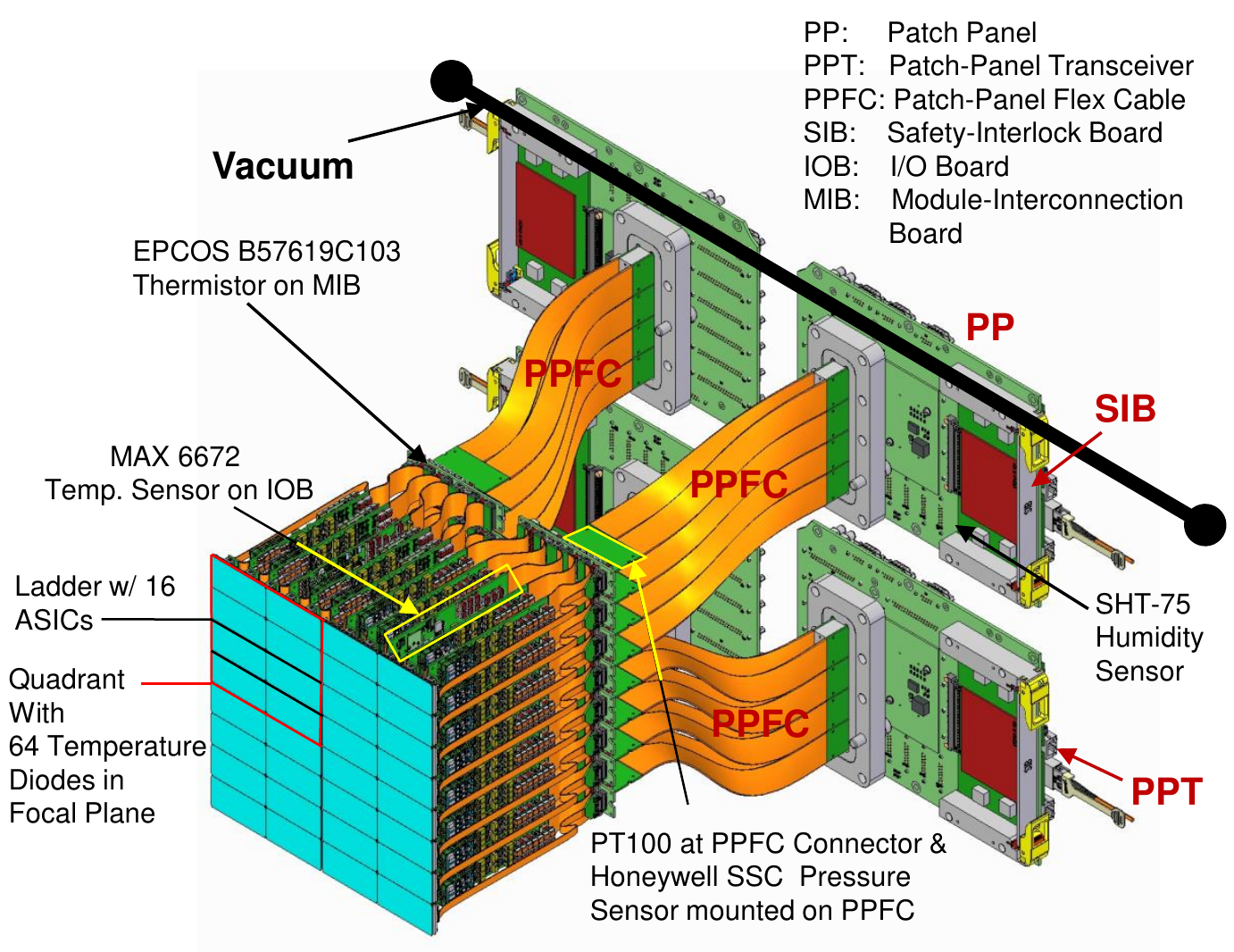}
\caption{DSSC mega-pixel X-ray imager with safety-interlock system}
\label{fig:capture1}
\vspace{-0.5cm}
\end{figure}

\IEEEPARstart{T}{he} DSSC X-ray imager is a novel mega-pixel camera based on DEPFET (Depleted P-Channel Field Effect Transistor) Sensor with Signal Compression (DSSC) \cite{IEEEhowto:kopka} optimized for single photon operation in the energy range between 500 eV and 6 keV at frame rates of up to 4.5 MFps for imaging experiments at the European XFEL \cite{IEEEhowto:kopxy}. As shown in Fig.\ref{fig:capture1}, the detector consists of a camera head connected to the patch panel (PP) electronics \cite{IEEEhowto:kopxk}. The camera head is divided into four quadrants, each consisting of four ladders. Each ladder comprises of two sensor chips of $128\times256$ pixel format, where a sensor is read out by eight ASICs. These ASICs are then read out by the I/O board (IOB) connected to the module interconnection board (MIB). The MIB is connected to the patch panel (PP) through patch panel flex cable (PPFC) allowing easy movement of the camera quadrant. Finally, the SIB and data collecting patch panel transceiver (PPT) (not visible in Fig.\ref{fig:capture1}) sit on the opposite side of the PP outside the vacuum. The quadrant data received at the PPT is sent for back-end storage. Each quadrant has its own set of power supplies for independent operation. A single cooling system is used for the whole camera.
 
The SIB protects each quadrant by identifying anomalies and probable hazardous situations in the detector chamber. During the start-up and operation phases, it takes necessary steps to bring and hold the detector into a secure state. It also allows a user-friendly operation by displaying the safety flag to the operator. The type and location of all the sensors placed in the vacuum chamber are presented in the Fig. \ref{fig:capture1}. In section II a detailed description of the hardware is presented followed by section III focusing on the software implementation and its working mechanism. In the end we show the test setup and results.        

\section{Hardware Description}

\begin{figure}[htbp]
\centering
\includegraphics[width=.5\textwidth,trim=0 0 0 0,clip]{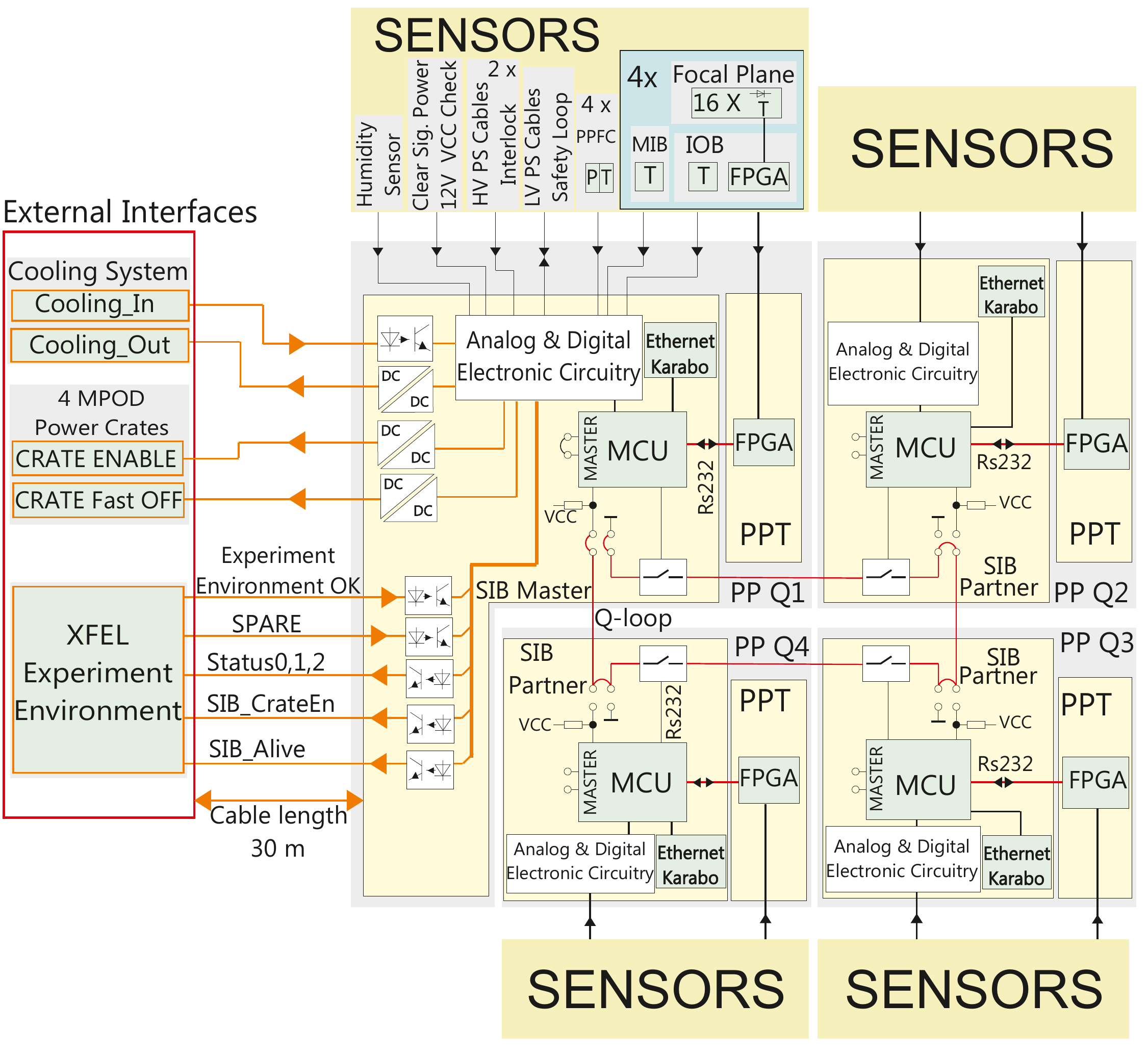}
\caption{SIB hardware block diagram}
\label{fig:blck}
\vspace{-0.2cm}
\end{figure}

Fig. \ref{fig:blck} shows a block diagram of all SIBs in a full mega-pixel camera setup with detailed hardware description. The SIB master performs all the functions similar to an SIB partner but with the exception, that SIB master can also monitor the signals from external interfaces. All the four SIBs communicate with each other through the quadrant loop (Q-Loop). Data is collected by each SIB from four temperature sensors of type MAX6672 on IOB, from four thermistors of type EPCOS-B57619 on MIB, from Honeywell SSCMNNN1 pressure sensors on PPFC and from temperature sensors Pt100 mounted on PPFC. A separate Pt100 temperature sensor is mounted on the cold trap (not shown). This cold trap device has the lowest temperature in the detector chamber. It condenses all the vapors being released in the vacuum chamber therefore preventing them from entering the vacuum pump which can get contaminated by condensation of these vapors. Temperatures of the Si-bare modules of the focal plane are monitored by diodes integrated on DEPFET sensor or on the readout ASIC (selectable). In both cases, digitized temperature values are provided per $64\times64$ pixel section and are sent through the PPT to the SIB via RS-232 interface. 
When the mains power switch is on, the power crate's auto-on channels turn on the SIB and PPT. Each SIB checks the power cable connection to the Wiener \cite{IEEEhowto:wener} LV-power crates by utilizing the safety-loop feature. Whereas in the ISEG \cite{IEEEhowto:iseg} HV-power modules there is in-built interlock functionality which prevents HV modules from being turned on in case of cable connection failure. The SIB monitors this failure by observing the current in the interlock lines via optocouplers.

\begin{figure}[htbp]
\centering
\includegraphics[width=.4\textwidth,trim=0 0 0 0,clip]{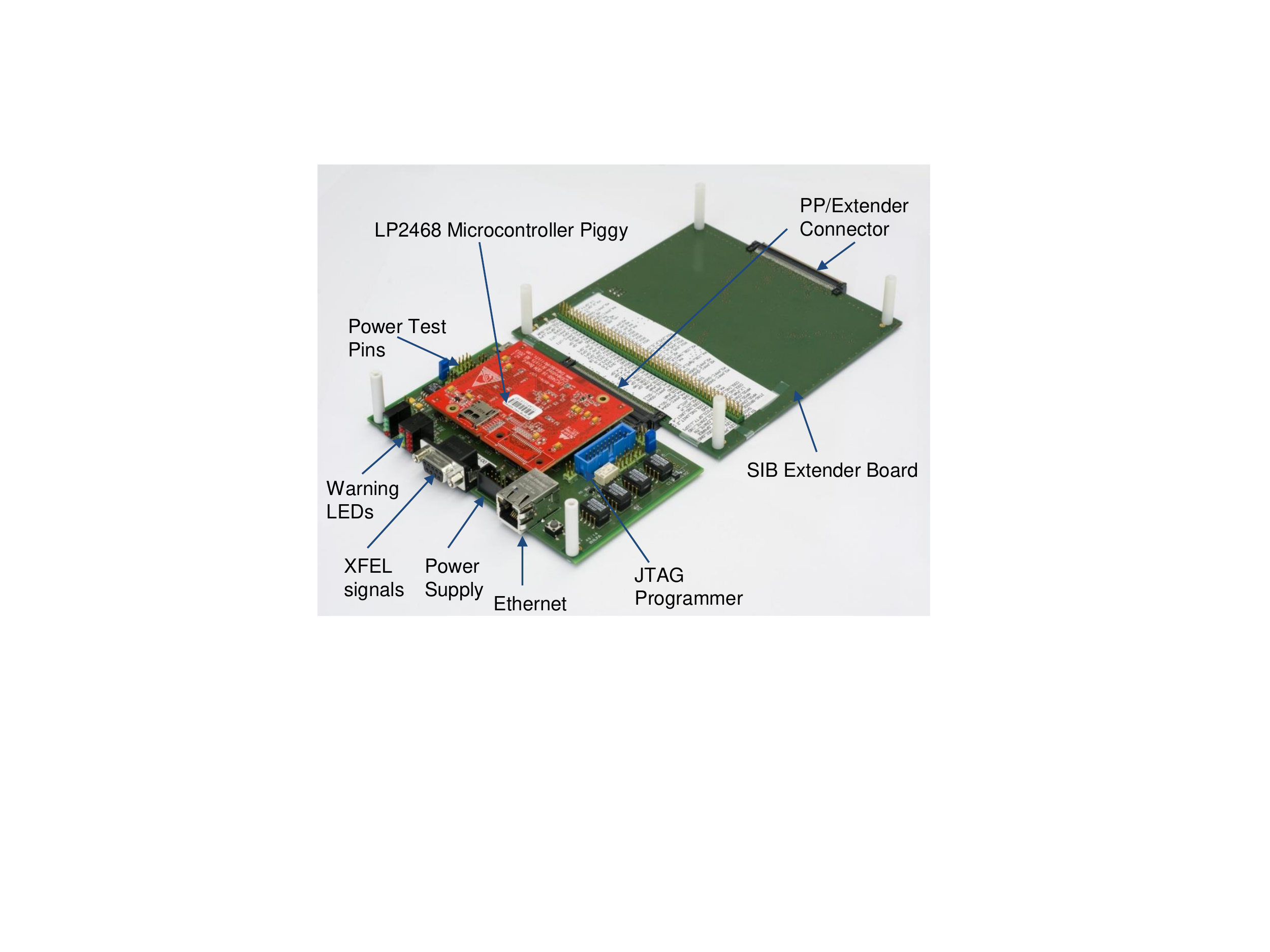}
\caption{Safety-interlock board with SIB-extender board}
\label{fig:sib}
\end{figure}
\begin{figure*}[htbp]
\centering
\includegraphics[width=.75\textwidth,trim=0 0 0 0,clip]{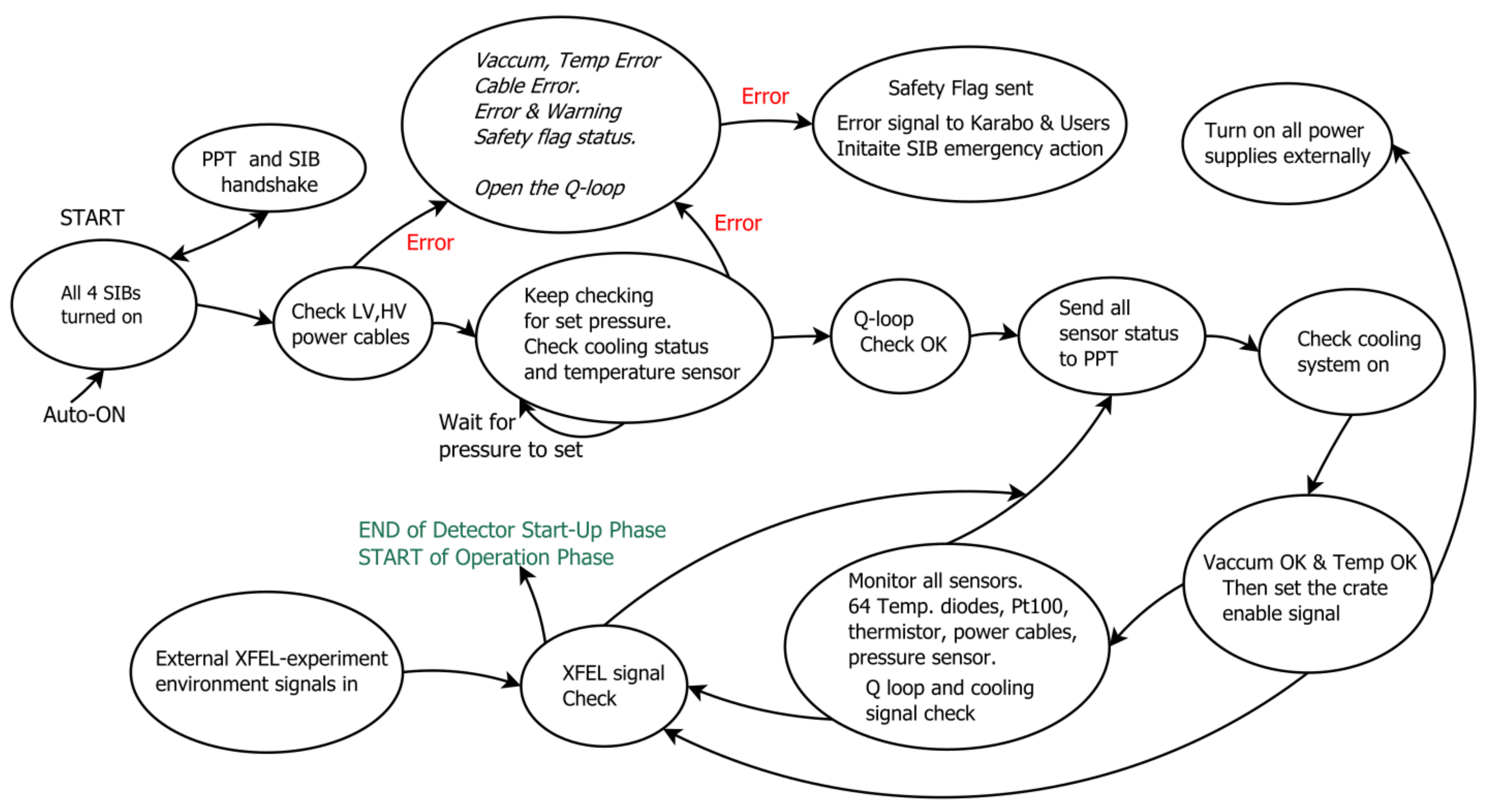}
\caption{Safety-interlock routine during the start-up phase}
\label{fig:sibrout}
\vspace{-0.4cm}
\end{figure*}
 
The special functionality of the SIB master is also shown in Fig. \ref{fig:blck}. Each crate has crate-enable and crate-fast off signals. The crate-enable signals to all the crates are connected in parallel and are operated once per camera by the SIB master. The crate-fast off has been implemented in a similar way to the crate-enable. Since this parallel connection is made externally it can be shortened for two or three quadrant operation. The SIB master interacts with the cooling system via two signals (Cooling\_In, Cooling\_Out) connected to the programmable logic control (PLC) Beckhoff \cite{IEEEhowto:beck} backend. Furthermore, two digital inputs and five outputs are provided for a direct communication between SIB master and XFEL-experiment environment. All signals to above mentioned external interfaces are isolated on board via optocouplers and isolated DC/DC converters. This blocks the noise and transients coming from external interfaces (via 30 m long cables). Since all the boards are designed identical in hardware, any of the four SIBs can become an SIB master by adding a jumper, for example in case of single quadrant operation. 

Fig. \ref{fig:sib} presents the developed SIB with all its components. It is an 8-layer PCB with the dimensions of 87x145x1.58 mm. The small piggy board with LPC2468-16/32 bit ARM7 controller running at 72 MHz is mounted on top. A separate power connector allows to perform tests using external power supplies in the absence of main power. The XFEL-experiment signals and cooling system signals are available via 15-pin connector. The error flags are also displayed on the LEDs as a live indicator to the end-user. The Ethernet jack, XFEL-signal connector and LEDs are mounted on the front-panel of SIB for robustness. The JTAG programmer lines allows fast hardware debugging during development phase. Later on, these signals will be connected to the Ethernet. In addition, an SIB-extender board was designed with signal test pins for integration tests.  

\section{Simplified Decision-Matrix Software in SIB}

Each SIB has a programmed decision-matrix which reads the sensor data and monitors external signals. Two pre-defined threshold values for each sensor is set in software indicating a warning state and an error state of the respective sensor. The failure and warning decisions are taken based on the sensor states as well as the Q-loop status. If one of the three SIB partners detects a critical sensor failure, it opens the Q-loop to notify the SIB master. For example, if a sensor read by SIB partner detects the pressure is above the set threshold, implying the vacuum is broken or leaking, Q-loop is opened and the SIB master is indicated. The SIB master sends a signal to deactivate the cooling system via Cooling\_Out line. Safety-flag is set with the sensor status of the individual quadrant by the corresponding SIB and sent to the back-end DAQ \cite{IEEEhowto:daq} along with image data of the bunch train every 100 ms. The Ethernet is implemented for receiving monitoring signals from the software control interface KARABO \cite{IEEEhowto:kopx}. It will also be used for remote software updates in expert mode.

In Fig. \ref{fig:sibrout}, different states of the decision-matrix are presented in a simplified state diagram. From the START state, SIB and PPT are alive before the camera-head sensors are powered on. The cable connection check starts and when it fails the SIB prevents other power supplies from being turned on. In the next step, the SIB waits for a certain pressure threshold (e.g. \sn{1}{-7} mbar) to be set externally by the vacuum pump, while it monitors the sensors. In the start-up phase, information from the vacuum pump (through KARABO) and the internal Honeywell pressure sensor together is used to take the safety decision. The internal pressure sensor can only detect a minimum pressure of 60 mbar which is sufficient to detect pressure leakages in the operation phase. Then, the Q-loop connection is checked by all SIBs. If all cables are connected and sensor states are within the expected ranges, the SIB waits for cooling system to be turned on and the Cooling\_In signal to be received. When a certain temperature level is reached and low pressure is achieved, only then the crate enable high signal is sent to the power crates, allowing all other power channels to be turned on externally. In case of an emergency, either of the following two methods can be used to turn off the power crates. The crate-enable signal can be pulled to low state turning off the crates by ramping down all power supplies \cite{IEEEhowto:iseg,IEEEhowto:wener} or the crate-fast off signal can be set to turn them off without ramping. Finally, the status input from the experimental environment is verified to ensure that the experimental setup is ready to start camera operation.

During the data-collection phase, SIB continuously monitors all sensor data. Each sensor has a \emph{warning} and an \emph{error} threshold. If the error threshold is breached emergency shutdown is initiated by SIB. In case of warning threshold violation, the end user decides to continue the experiment or initiate a downtime. 

\begin{figure}[htbp]
\centering
\includegraphics[width=.5\textwidth,trim=0 0 0 0,clip]{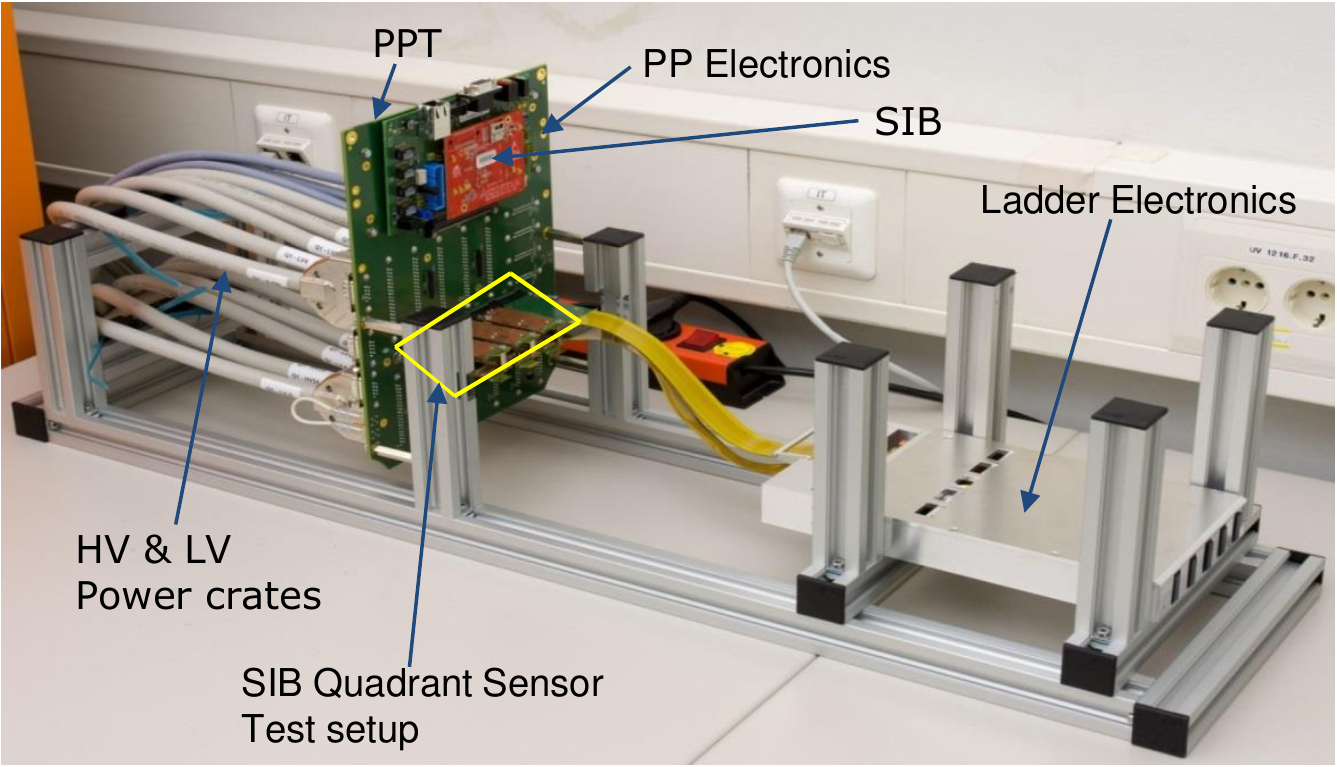}
\centering
\caption{Full tests at quadrant-test stand}
\label{fig:qts}
\end{figure}

\begin{figure}[htbp]
\centering
\includegraphics[width=.5\textwidth,trim=0 0 0 0,clip]{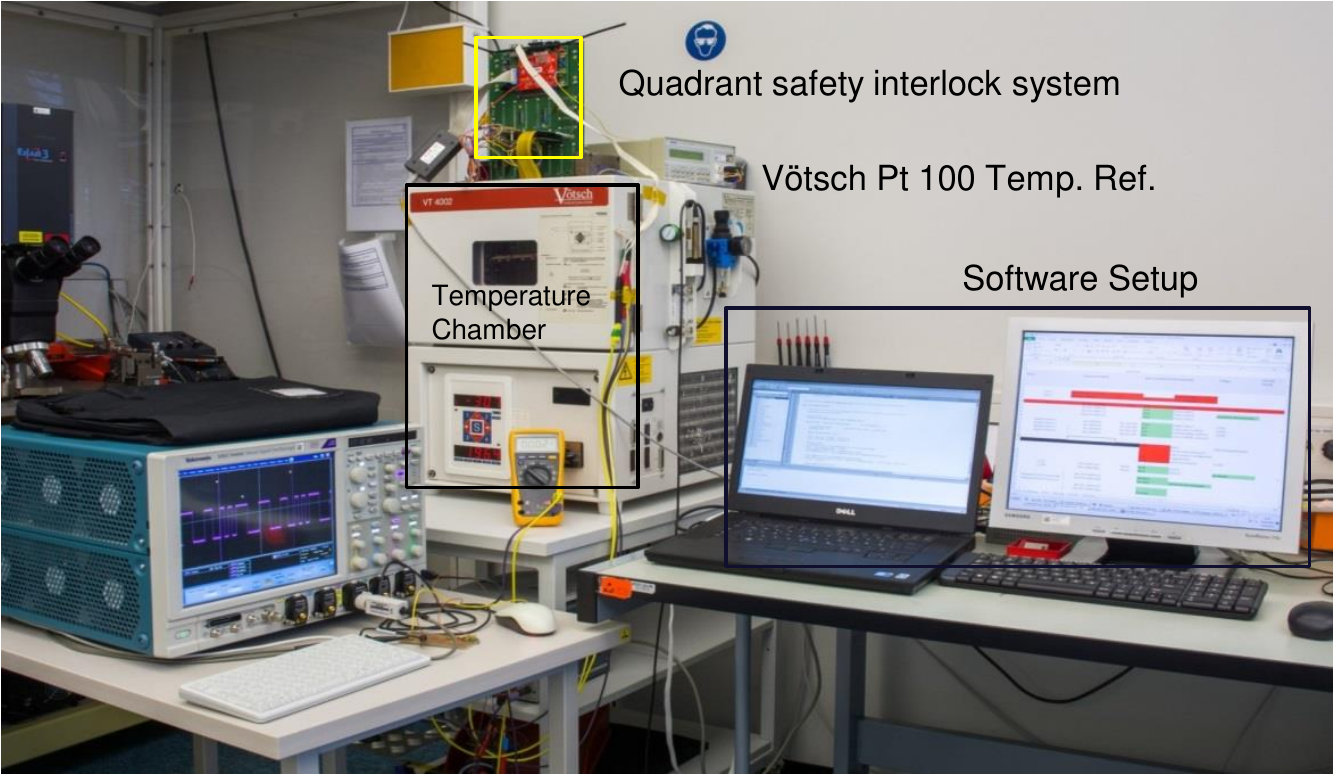}
\centering
\caption{Ladder tests at V\"otsch temperature chamber}
\label{fig:ladtest}
\vspace{-0.4cm}

\end{figure}

\section{Measurement Setup and Test Results}

Fig. \ref{fig:qts} shows the quadrant-test stand (QTS) with a fully equipped power crates including 30 m long cables connected to the PP and one ladder electronics. Additional three test PCBs were developed with the remaining sensors for full-quadrant tests. They were plugged into the remaining PPFC connectors. The crate power settings for PPT and SIB was tested before plugging in the boards at the QTS. The software was downloaded on the micro-controller via JTAG interface. All the sensor data from the pressure and temperature sensors were read out successfully. The RS232 communication between the SIB and PPT was established and all the sixty-four temperature diode were read out from the PPT as dummy data. Furthermore, LV safety loop and HV interlock signals were tested from the crate-controller. The cooling and XFEL-experiment signals were checked with the Beckhoff EL2124 system. The crate-enable and crate fast-off signals were monitored at the connectors on the PP. The Q-loop was tested by closing the connection between input and output on PP. 

\begin{figure}[htbp]
\begin{center}
\includegraphics[width=.45\textwidth,trim=0 0 0 0,clip]{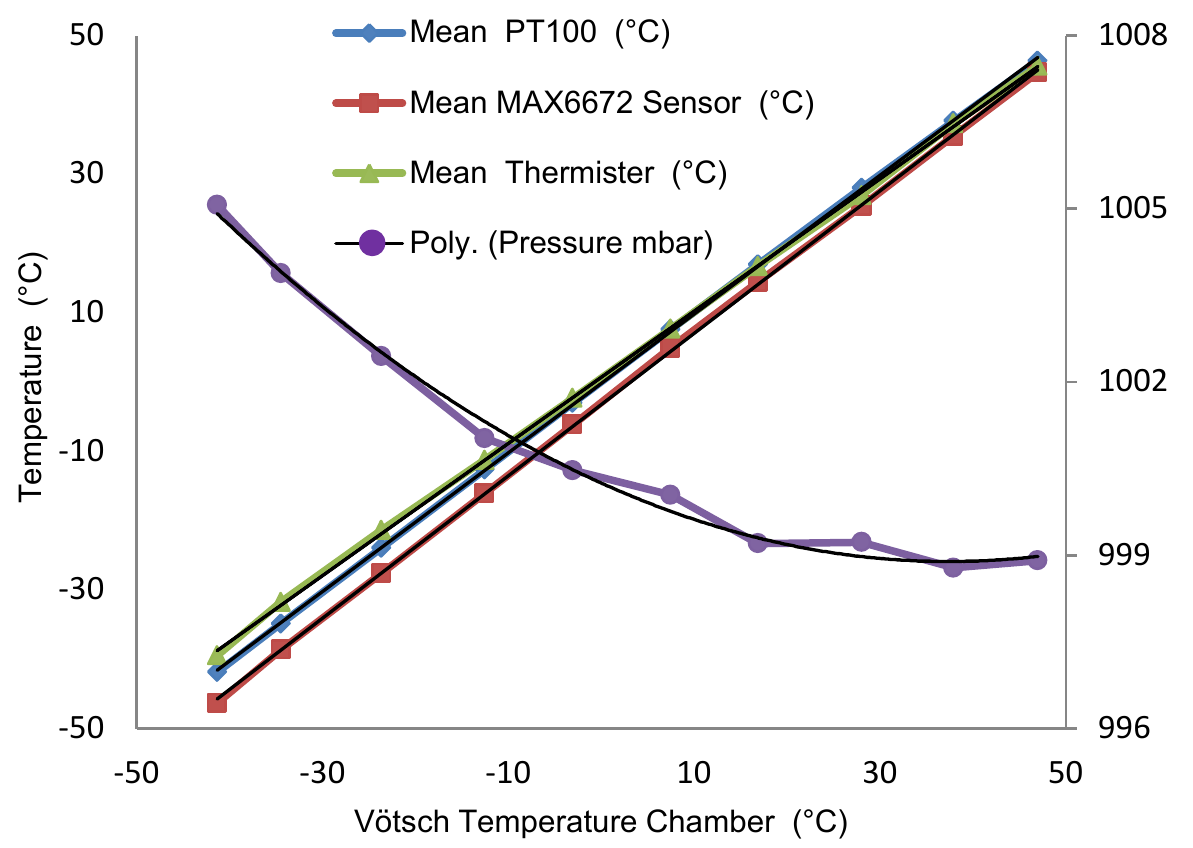}
\includegraphics[width=.45\textwidth,trim=0 0 0 0,clip]{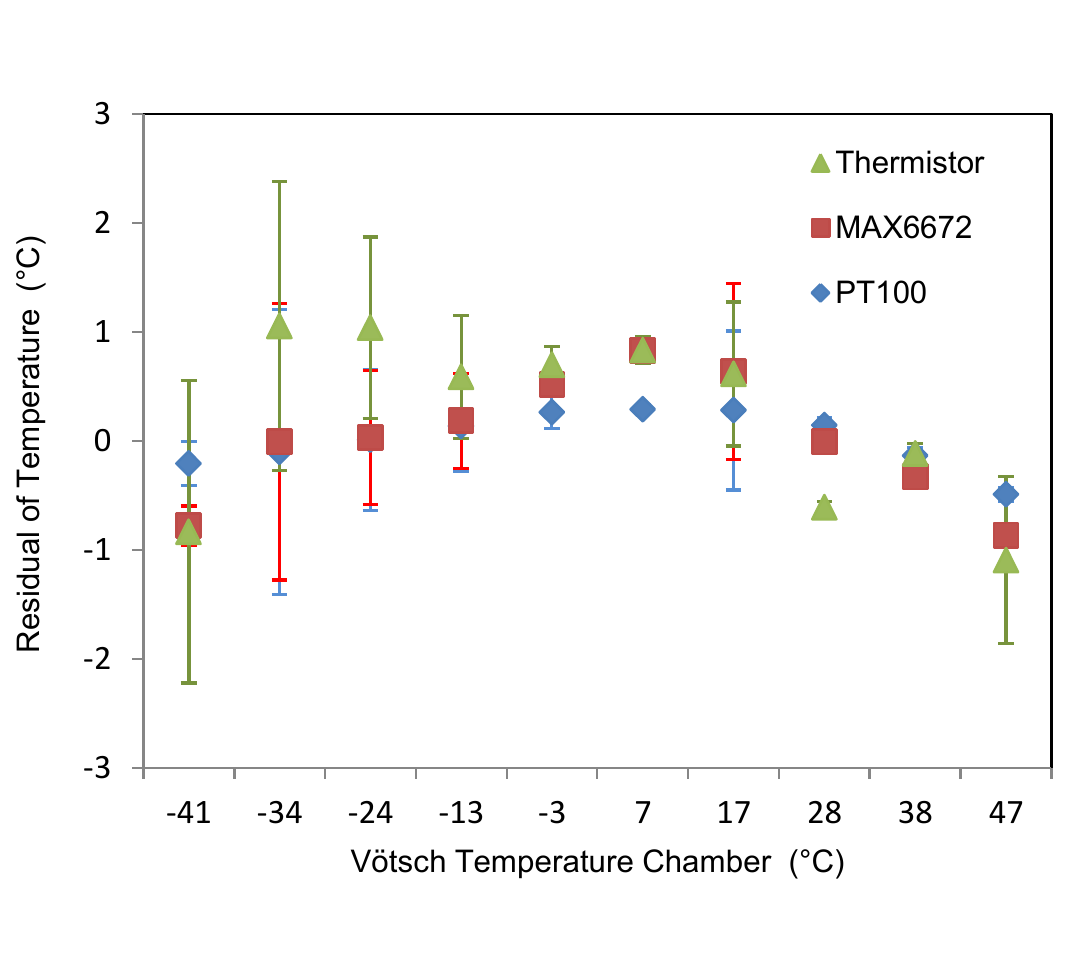}
\includegraphics[width=.43\textwidth,trim=0 0 0 0,clip]{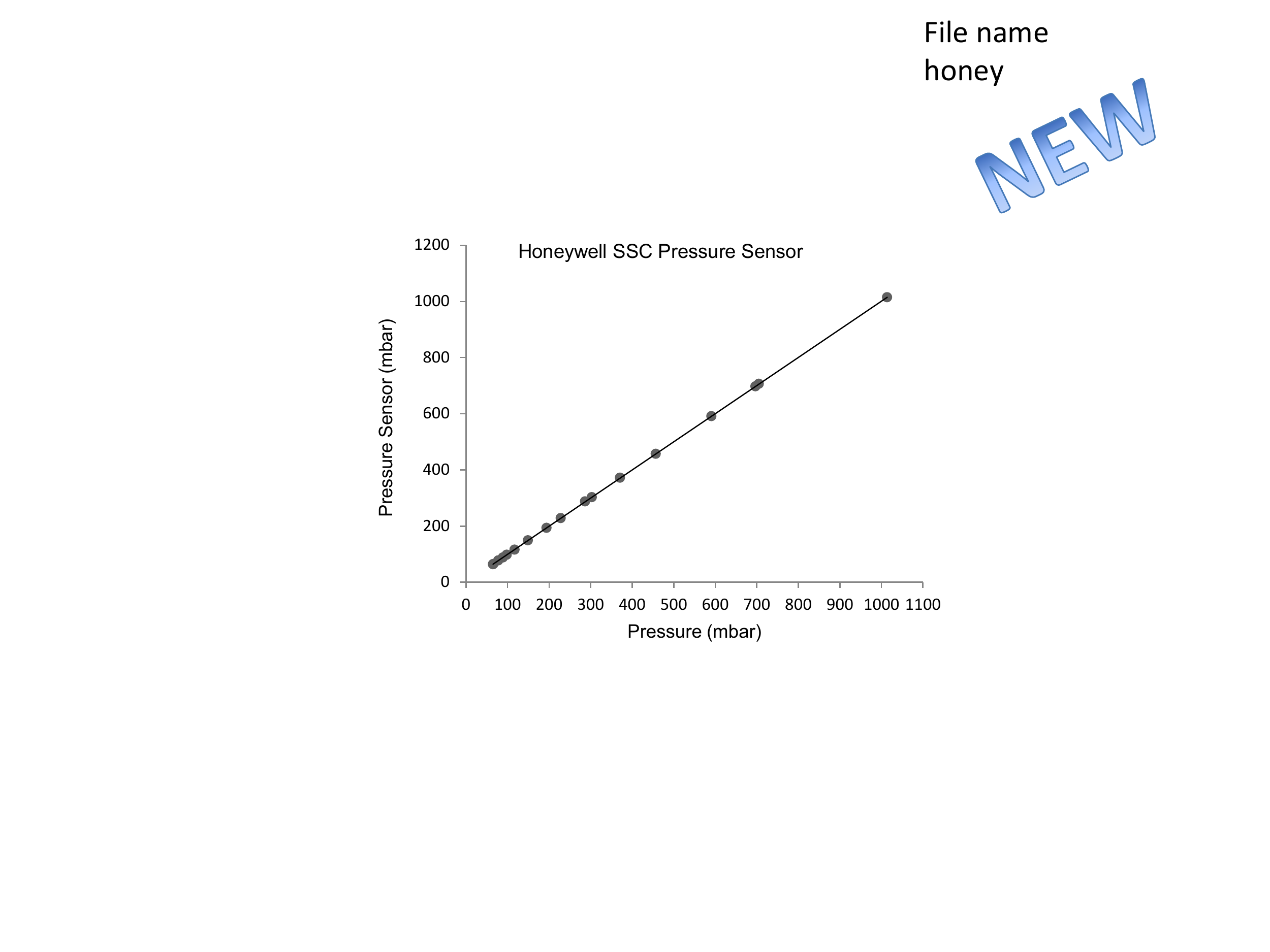}
\caption{Sensor calibration graphs}
\label{fig:capsensx}
\vspace*{-0.8cm}
\end{center}
\end{figure}

For the calibration tests, measurement of the temperature sensors was performed in a temperature controlled test chamber VT4002 (V\"otsch GmbH) Fig. \ref{fig:ladtest}. The temperature was varied in the range of $-40^ {\circ}$C to $+50^{\circ}$C. The temperature sensors were calibrated with respect to the Pt100 reference sensor of the test-chamber. The achieved accuracy of the Pt100 on PPFC and NTC thermistor sensors in all the four ladders was $\pm 1^ {\circ}$C. The MAX 6672 sensor has a low precision of $\pm 5^ {\circ}$C (especially at lower temperatures of $-40^ {\circ}$C) as mentioned in the datasheet. It is the one of the worst performing sensors of the system. The pressure sensors recorded an accuracy of $\pm 1$mbar Fig. \ref{fig:capsensx}. 

In the end, timing analysis was done with the quadrant software module at the QTS to determine the time required by the micro controller to read all sensors, calculate the corresponding sensor values and send them to PPT. In the quadrant level tests, the four universal transducer interface (UTI) devices connected to the PT100 were the slowest sensors (128 ms). The smallest acquisition time was for the four pressure sensors (3 ms). The overall processing time for all sensors in each SIB was 500 ms including the data transfer to the backend. Hence it can be concluded that the full sensor data will be sent via PPT every 600 ms. And only the safety flag will be sent every 100 ms. When a sensor reaches error threshold in the decision matrix further execution will be stopped and emergency action will be initiated by the SIB.

\section{Conclusion}
The safety-interlock system was developed and tested. All the above tests were performed at the production-ready detector setup. The probable critical and failure conditions during the detector operation were simulated. The first software dry run on SIB together with the external devices was done successfully.

\end{document}